\documentstyle[12pt]{article}
\begin{document}

{\bf QCD SUM RULES AND HADRONIC RADIAL EXCITATIONS}

{\bf ~~~~~~~~~~~~~~~~~~~~~~~~~~~~~~OF LIGHT MESONS}

\begin{center}

{\bf A.L. Kataev}\\

{\em Institute for Nuclear Research of the Academy of
Sciences of Russia,\\ 117312 Moscow, Russia}

\end{center}

It is well known that the QCD sum rules method was formulated
in its classical form in Ref.\cite{ShVZ},
where it was proposed to relate the properties of
the ground states of hadrons to the number of the nonperturbative
fundamental QCD parameters, namely quark and gluon
condensates $<\overline{\psi}\psi>$, $<\alpha_s G_{\mu\nu}^2>$ ,
using the operator product expansion method and the realization
of the global duality hypothesis, implemented through the Borel-type
dispersion sum rules.
However, the application of the procedure of the
Borelezation can   lead to the suppression of the
radial excitations of hadronic ground states contributing
to the physical spectral density of sum rules.

In this talk we remind that it is possible to study the
properties of these radial excitations using the
finite-energy QCD sum rules \cite{KTCh}, which is the
QCD generalization of the method  of the dual sum rules,
formulated in Ref.\cite{LST} to relate high-energy Redge
behavior of the hadron-hadron cross-sections with the
characteristics of low-lying resonances. This
approach, developed in Ref.\cite{KP} for the case of
$\rho$, $\phi$, $J/\Psi$ and $\Upsilon$ channels, was further
used for the estimation of the properties of radial excitations
of $\pi$-meson in Ref.\cite{KKP} and $K$-meson in Ref.\cite{GKL},
where the predictions for the masses and leptonic decay constants
of the radial excitations of the light scalar mesons $a_0(980)$,
and $K_0^*(1430)$ were also obtained in the cases when both
$K_0^*(1430)$ and $a_0(980)$-mesons were considered as the
quark-antiquark bound states.

It is stressed that in the $\rho$-meson channel the derived
in Ref.\cite{KP} predictions
\begin{equation}
m_{\rho}^2(n)=m_{\rho}^2\times(1+2n)~~~,~~~\Gamma_{n}^{e^+e^-}=
\frac{\alpha^2 m_{\rho}^2}{3\pi(2n+1)}
\end{equation}
gave one of the first arguments in favor of the fact
that the particle $\rho(1700)$
should be considered as the {\bf{second}} radial excitation of
of $\rho$-meson and that  its {\bf{first}} radial excitation should have
the mass of over $1300~MeV$. This prediction  is now confirmed
at CERN
by the Crystal Barrel Collaboration through the
observation  of the first radial excitation of $\rho$-meson
with the mass $m_{\rho^{'}}=1411\pm 10~ (stat)\pm~10~{syst}~MeV$ in
the reaction $\overline{p}n\rightarrow\pi^-\pi^0\pi^0$
at LEAR \cite{CrBar}.

In the case of $\pi$- and $K$-meson channels the chiral limit FESR prediction
is \cite{KKP}
\begin{equation}
m_{PS}^2(n)=m_{{PS}^{'}}^2\times n~~~,~~~f_{PS}(n)=2\sqrt{2}
\bigg(\frac{m_{PS}^2}{m_{{PS}^{'}}m_{PS}(n)}\bigg)f_{PS}
\end{equation}
where $n\geq 1$, $PS$ means the abbreviation
for the pseudoscalar mesons $\pi$ and $K$
and
$m_{\pi^{'}}\approx 1.24~GeV$ was measured some time
ago at Protvino \cite{DMB}. The following from Eq.(2) prediction
$m_{\pi}(2)\approx 1.75~GeV$ is in good agreement with the observation
of another $I^G J^P L =1^- 0^{-}S$ resonance at $1.77 \pm 0.03~GeV$
\cite{DMB} which could correspond to a second radial excitation of
the pion $\pi^{''}$. This experimental result was confirmed recently
also at Protvino by  VES Collaboration \cite{VES}. It should be
mentioned, however, that the experimental data of the VES
collaboration do not exclude the possibility that the corresponding
resonance with the mass $M=1786\pm 7(stat)\pm30(syst)~MeV$
has exotic (hybrid-like) nature. In view of this it can be of
real interest to study carefully the enhancement in the
$\rho\pi$-channel of the system $\pi^+\pi^-\pi^-$ near the
invariant mass of order $2.1-2.4~GeV$ (see Fig. 1. (c) of Ref.\cite{VES}),
where due to the FESR prediction of Ref.\cite{KKP} the third radial excitation
$\pi^{'''}$ of the $\pi$-meson can manifest itself.

In the scalar channel the predictions of the FESR read \cite{GKL}
\begin{equation}
m_{S}^2(n)=m_S^2\times(n+1)~~~,~~~f_S^2(n)=\frac{1}{n+1}f_S^2~~~.
\end{equation}
The  strange candidate to the scalar (S) ground state
is the $K^{*}(1430)$-meson.
In view of the predictions of Eq.(3)  it is reasonable to expect,
that its first radial excitation can have the mass of over
$2~GeV$. This expectation is in good agreement
with the experimental indications
to the existence of its first radial excitation, namely
$K^{*}(1950)$-meson \cite{PD}.

The situation in the non-strange scalar sector is
more intriguing. Despite the fact that there are the
studies of the possibilities that the lightest non-strange
scalar meson $a_0(980)$ can be the four-quark state or
the $\overline{K}K$-molecule (see e.g. Ref.\cite{A}), there are also
the arguments in favor of the standard $\overline{q}q$-structure
of the $a_0(980)$-meson (for one of
the latest related works see Ref. \cite{AS}).
The FESR considerations of  Ref.\cite{GKL}, which were made
in this case, are giving
the predictions of the mass of radial excitation of the $a_0(980)$-meson,
namely $m_{a_0^{'}}\approx 1.4~GeV$ \cite{GKL}.
It is  interesting to note that not long ago the
Crystal Barrel Collaboration at CERN have found the signal
of the non-strange scalar particle $a_0(1450)$ in the reaction
$\overline{p}p\rightarrow \pi^0\pi^0\eta$ \cite{a0}.

It is also worth to add several words
about  one more QCD sum rules predictable
characteristic of light mesons, which is directly
related to the experimental data, obtained at
Protvino. This characteristic is the ratio $\gamma=F_A/F_V$
of the axial and vector form-factors of radiative decay of charged
pseudoscalar meson $\pi^+\rightarrow e^+\nu\gamma$, which provides
the important information about the structure of weak hadronic currents
(for the review see Ref.\cite{PR}).
The application of the
three-point function QCD sum rules
and of the generalized operator product expansion, formulated in
the works of Ref.\cite{GOPE}, resulted in the following value
of this parameter \cite{KOP}:
\begin{equation}
\gamma=0.51 \pm 0.13~~~ (theory,~ QCD~sum~rules)~.
\end{equation}
It is in perfect agreement with the outcome of the latest
most precise measurement of this parameter, obtained by
the Istra collaboration at the Protvino accelerator \cite{Istra}:
\begin{equation}
\gamma=0.41 \pm 0.23~~~ (experiment)~.
\end{equation}

This note is prepared within
the framework of the program of the research of the Project
N 96-01-01860 of the
Russian Fund for Fundamental Research.

\end{document}